\documentclass[11pt]{article}
\usepackage{amsmath,amsfonts,times,pictex,epsf}

\renewcommand{\theequation}{\thesection.\arabic{equation}}

\newcommand{\appsection}{
\setcounter{equation}{0}
\setcounter{section}{1}
\section*{Appendix}
\renewcommand{\theequation}{\Alph{section}-\arabic{equation}}}

\newcommand{\eps}{\varepsilon}
\newcommand{\hcap}{\text{hcap}}
\newcommand{\One}{{1 \!\! 1}}

\newcommand{\cD}{\mathcal{D}}
\newcommand{\cF}{\mathcal{F}}
\newcommand{\cL}{\mathcal{L}}
\newcommand{\cO}{\mathcal{O}}
\newcommand{\mE}{\mathbb{E}}
\newcommand{\mH}{\mathbb{H}}

\begin{document}


\begin{center}
{\Large {\bf On Multiple Schramm-Loewner Evolutions}} \\[5pt]
\end{center}
\vskip 1.8cm
\centerline{K.~Graham\footnote{e-mail: {\tt kgraham@physik.fu-berlin.de}}
}
\vskip 0.6cm
\centerline{17th November 2005 }
\centerline{${}^1$\sl Freie Universit\"at Berlin, Department of Physics,}
\centerline{\sl Arnimallee 14, 14195 Berlin, Germany}\vskip 0.9cm
\begin{abstract}
\vskip0.15cm
\noindent
In this note we consider the ansatz for Multiple Schramm-Loewner Evolutions (SLEs) proposed by Bauer, Bernard and Kyt\"ol\"a from a more probabilistic point of view.  Here we show their ansatz is a consequence of conformal invariance, reparameterisation invariance and a notion of absolute continuity.  In so doing we demonstrate that it is only consistent to grow multiple SLEs if their $\kappa$ parameters are related by $\kappa_i = \kappa_j$ or $\kappa_i = \tfrac{16}{\kappa_j}$. 
\end{abstract}
\setcounter{footnote}{0}
\def\thefootnote{\fnsymbol{footnote}}

\section{Introduction}

Schramm (or Stochastic) Loewner Evolutions (SLEs) are a powerful tool
to describe the continuum limit of two-dimensional interfaces in
statistical mechanics at criticality \cite{Sch9904,RohSch0106}\footnote{For an
introduction to this field we suggest \cite{Wer0303,KagNie0312}.
There was also a useful preprint by Lawler which is now a book
\cite{book:Law}.}.  Since such statistical mechanic models are also
expected to have a conformal field theory interpretation, we are
naturally led to understand the connections between conformal field
theory (CFT) and SLE \cite{BauBer0210,FriWer0301}.

In the case were configurations involve just one interface, the
connections are well understood largely due to the work of Bauer and
Bernard \cite{BauBer0210,BauBer0305,BauBer0401}.  However, to
understand if there is a role for such CFT notions as fusion and
conformal blocks it is necessary to consider configurations involving
many interfaces \cite{Car0310,Dub0411}.

The question what is the correct SLE description of multiple
interfaces consistent with conformal symmetry has been addressed in a
paper by Bauer, Bernard and Kyt\"ol\"a (BBK) \cite{BauBerKyt0503}
wherein the authors couple CFTs to multiple SLEs and find the
conditions for certain objects to be martingales.  From this they
provide an ansatz conjectured to describe multiple SLEs consistent
with conformal invariance.  While this procedure is justified from the
statistical mechanic point of view, it is not clear how to interpret
elements of their ansatz in probability theory.

In this note we consider multiple SLEs from a more probabilistic
point of view.  By assuming conformal invariance, reparameterisation
invariance of the curves and a notion of absolute continuity we
rederive the BBK ansatz.  Along the way we demonstrate that it is only
consistent grow to multiple SLEs if their $\kappa$ parameters are
related by $\kappa_i = \kappa_j$ or $\kappa_i = \tfrac{16}{\kappa_j}$,
hence we find realisations of both the conformal field theory fields
$\phi_{1,2}$ and $\phi_{2,1}$: all the building blocks needed to
create general fields $\phi_{r,s}$ from the Kacs table.  This
condition can be restated as saying multiple SLEs can only be grown
consistently if they all have the same central charge.

The plan of the paper is as follows:  First we introduce multiple SLEs
and the ansatz of BBK, pointing out some of the questions answered in
later sections.  We then discuss an application of absolute
continuity by looking at the connection between single and multiple
SLEs.  Section \ref{sec:conformal} studies the requirement of
conformal invariance on multiple SLEs, while section
\ref{sec:commute} considers reparameterisation invariance following
the work of Dub\'edat \cite{Dub0411}\footnote{We note some overlap
between the work presented here and the second version of
\cite{Dub0411}.}.  In section \ref{sec:timestwo} we reconsider
reparameterisation invariance using a different technique and observe
the same results.  We end with our conclusions.

\section{Multiple SLEs and the Ansatz of Bauer, Bernard and
Kyt\"ol\"a}

The primary object of study in this paper is a space of $m$ curves in the closure of the upper half plane (UHP) that start
and end on the real  line plus infinity.  These curves may have self
intersections and mutual intersections but no crossings.  For our
purposes it is more helpful to think of $n$-curves starting on the
real axis at points $x_0^i$: $2m{-}n$ of these curves go to infinity
while the remaining $n{-}m$ pair-up to form the total of $m$ complete
curves.

To study this space and probability measures upon it we follow the
idea of Schramm \cite{Sch9904} to use Loewner evolutions to formulate
an equivalent description of the curves as real valued functions.
Then the measures on curves lift to measures on a space of real valued
functions.  The goal of this paper is to study how properties of the
curve measure (such as conformal invariance and reparameterisation
invariance) translate into properties of the measure on driving
functions.

The Loewner evolution map for multiple curves is the solution to,
\begin{align} 
  \dot{G}_t(z) = \sum_{i = 1}^n \frac{2 a_t^i }{G_t(z) -x_t^i}  \; ,  
\qquad 
  G_0(z) = z \; .   \label{eq:defn:G}
\end{align}  
This equation is well defined up to some explosion time
$\tau_z = \inf\{ t : G_t(z) \in \{ x_t^1 , \ldots , x_t^n \} \}$.  The
set $K_t = \{ z \in \mH : \tau_z \le t \}$ is called the hull and is
such that $G_t : \mH / K_t \to \mH$, $\mH$ denoting the upper half
plane.  One can recover the curves $\gamma_t^i$ from the $x_t^i$ and
$a_t^i$ via $\gamma_t^i = \lim_{\eps \to 0} G_t^{-1}(x_t^i {+} i \eps)$, then the hull is the
component of the set $\mH / \cup_i \gamma_t^i$ connected to infinity:
here we use the notation $\gamma_t^i$ to denote both the location of
the tip of the curve $i$ at time $t$ and the set $\{ \gamma_s^i : s
\le t\}$, we do not expect any confusion.

There are of course many maps which conformally map $\mH / K_t \to
\mH$.  However, as a solution to \eqref{eq:defn:G}, $G_t$ is
automatically hydrodynamically normalised, that is to say it is the
unique conformal map $\mH / K_t \to \mH$ such that $\lim_{z \to
\infty} G_t(z) - z = 0$.  In this paper we will meet a lot of
conformal maps mapping $\mH / A \to \mH$ for some set $A \subset \mH$, it
will always be implicit that these maps take the component of $\mH / A$ connected to infinity onto $\mH$ and are hydrodynamically normalised.

A natural quantity in Loewner evolutions is the upper half plane
capacity, $\hcap$, which for a hull $K_t$ is defined via the expansion,
\begin{align} 
  G_t(z) = z + \frac{2 \hcap[ K_t ]}{z} + \cO(z^{-2}) \; ,
\end{align} 
(note the factor of $2$ in this definition) then if the curves $\gamma_t^i=\gamma_{t^i(t)}^i$ are individually parameterised by $t^i(t)$, $K_t = K_{t^1\!(t), \ldots , t^n\!(t)}$ we find from \eqref{eq:defn:G},
\begin{align} 
  a_t^i = \frac{dt^i(t)}{dt} \frac{\partial}{\partial t^i} \hcap[ K_{t^1\!(t), \ldots , t^n\!(t)} ] \; .
\end{align} 
It is important to note that the pair $(x_t^i, a_t^i)$
encode the curves and their parameterisation: changing the
parameterisation changes $x_t^i$ and $a_t^i$ in some complicated way.

In the above, we have started with the functions $(x_t^i,a_t^i)$ and
constructed curves with a parameterisation.  To go the other way, one
takes the maps $G_t$ associated with the curves and notes that they
satisfy an integral equation generalising \eqref{eq:defn:G} which
defines functions $x_t^i$ and measures $a_t^i dt= \hcap[ K_{t^1\!(t),
\ldots, t^i\!(t{+}dt), \ldots , t^n\!(t)} ]$.   To write
\eqref{eq:defn:G} we must assume some absolute continuity for these
measures.  We will do this for simplicity.

\subsection{The Ansatz of Bauer, Bernard and Kyt\"ol\"a}

In the case of SLE for a single curve, it was shown by Schramm
\cite{Sch9904} that conformal invariance implies the driving function
$x_t$ should be a continuous Markov process which, in a particular
parameterisation, has independent increments.  This requires
 $x_t = \sqrt{\kappa} B_t$ for some
standard Brownian motion $\langle B,B \rangle_t = t$.  By a time
change, in a general parameterisation $x_t$ is a continuous martingale
with quadratic variation $ \langle x,x \rangle_t = \kappa \, \hcap [
K_t ]$.

\begin{figure}
\refstepcounter{figure}
\label{fig:Gs}
\begin{center}
\begin{tabular}{c}
\setlength{\unitlength}{1cm}
\begin{picture}(12,4)
  \put(0,2){
     \epsfxsize 5\unitlength
     \epsfbox{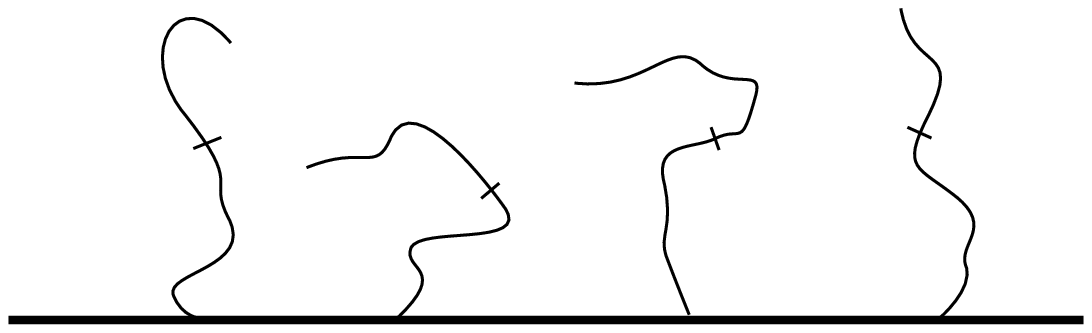}}
  \put(7,2){
     \epsfxsize 5\unitlength
     \epsfbox{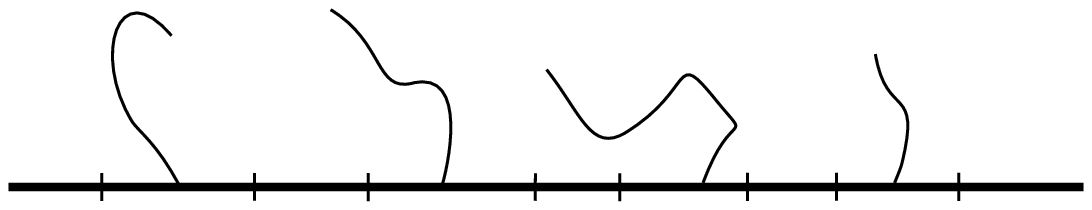}}
  \put(3.5,0){
     \epsfxsize 5\unitlength
     \epsfbox{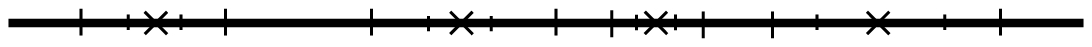}} 
  \put(5.5,2.5){
     \epsfxsize 1\unitlength
     \epsfbox{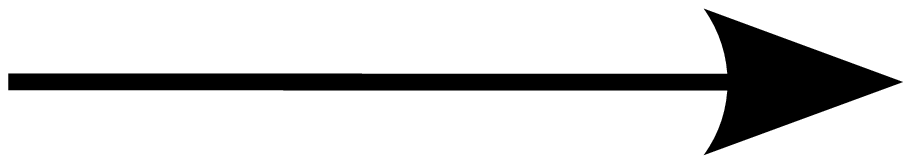}}
  \put(8.75,0.75){
     \epsfysize 1\unitlength
     \epsfbox{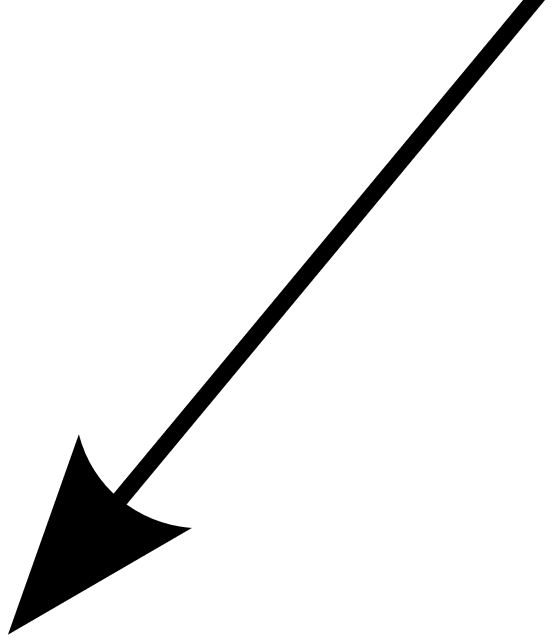}}
  \put(2.5,0.75){
     \epsfysize 1\unitlength
     \epsfbox{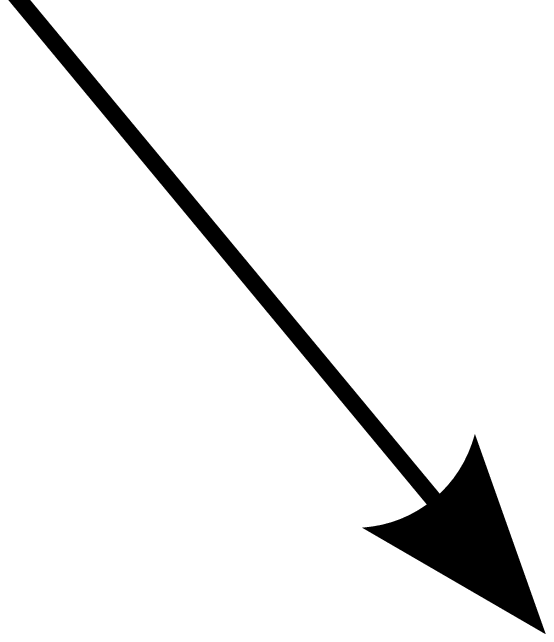}}
  \put(5.8,2.85){ \makebox(0,0)[l]{$G_s$} }
  \put(2.3,1.1){ \makebox(0,0)[l]{$G_t$} }
  \put(9.4,1.1){ \makebox(0,0)[l]{$G_{t,s}$} }
\end{picture}
\vspace{0.5cm} \\
\parbox{12cm}{
\small\raggedright
{\em \begin{center} Figure \ref{fig:Gs} : The maps $G_t$ and $G_s$ acting on curves $\gamma_t^i$. \\
In the left hand diagram, the lines represent the curves up to time $t$, while the dashes across the curves represent the positions of the tips at time $s$.  In the right hand diagram we have used dashes to mark the images of the points $\gamma_0^i$ under $G_s$. \end{center}}}
\end{tabular}
\end{center}
\end{figure}

For multiple SLEs, the same argument shows the driving functions
$x_t^i$ should be a continuous Markov process.  Indeed, consider
distribution of $n$-multiple SLE curves in the upper half plane and
consider two Loewner maps $G_t$ and $G_s$, $s<t$ (see figure
\ref{fig:Gs}).  Note that $G_{t,s} = G_t \circ G_s^{-1}$ is the
solution to,
\begin{align}
  \frac{ dG_{t,s}(z) }{ dt } = \sum_i \frac{ 2 a_t^i}{G_{t,s}(z) -
    x_t^i} \; ,
  \qquad
  G_{s,s}(z) = z \; .
\end{align}
By conformal invariance, the distribution of $G_{t,s}$ given $x_s^i$
should be the same as a Loewner evolution starting from $x_s^i$ and
parameterised by $a_t^i$ in the same way.  In particular, this distribution should
be independent of $x_r^i$ for $r<s$.  As the distribution of $G_{t,s}$ is
encoded in the distribution of $x_t^i$, this implies the distribution
of $x_t^i$ given $x_s^i$ should be independent of $x_r^i$, $r<s$ and so $x_t^i$ is a Markov process.

However, we need to do more work to extract all the consequences of conformal invariance.  In \cite{BauBerKyt0503}, BBK studied multiple SLEs by coupling them to conformal field theory.  By arguing that certain CFT quantities should be martingales, they obtained the following stochastic differential equation which the driving functions $x_t$ consistent with conformal symmetry should satisfy,
\begin{align}
  dx_t^i = dM_t^i 
   + \kappa_i a_t^i \frac{\partial}{\partial x_t^i} \log Z[x_t] dt
   + \sum_{k \ne i} \frac{2 a_t^k}{x_t^i - x_t^k} dt \; , 
  \label{eq:xeq}
\end{align}
where $M_t^i$ are independent continuous martingales with quadratic variation $d\langle M^i,M^j \rangle_t = \kappa_i a_t^i \delta_{ij} dt$, the function $Z[x]$ transforms as a tensor under M\"obius transformations and it satisfies the ``null vector equations'',
\begin{align}
   0 = \frac{\kappa_i}{2} \frac{\partial^2 Z[x]}{\partial x_i^2} 
   - \sum_{k \ne i} \frac{2}{x_i - x_k} \frac{\partial Z[x]}{\partial x_k} 
   - \sum_{k \ne i} \frac{2 h_k}{(x_i - x_k)^2} Z[x] \; , 
  \label{eq:nve1}
\end{align}
In their paper, BBK go on to conjecture how the space of solutions to \eqref{eq:nve1} represent the different ways of joining the $n$-multiple SLEs to create $m$ simple curves.  We refer the reader to BBK for details.\\

In this note, we will concentrate on understanding the origin of \eqref{eq:xeq} and \eqref{eq:nve1} without reference to CFT.  We will show that the form of \eqref{eq:xeq} and the null vector equation follow from conformal and reparameterisation invariance.  Some other questions will also be answered:

1. Why does the quadratic variation of the driving martingales have the form $d\langle M^i,M^j \rangle_t = \kappa a_t^i \delta_{ij} dt$?  In BBK, they argue that the driving functions should ``grow independently of each other on short time scales'', but what does this phrase mean?  In the next section we answer this question using a notion of absolute continuity.

2. Do all the martingales have the same $\kappa$ parameter?  In section \ref{sec:commute} we will see that this assumption maybe relaxed to $\kappa_i \in \{ \kappa, \tfrac{16}\kappa \}$ and that the restriction is due to reparameterisation invariance.

\subsection{Girsanov's theorem}

Girsanov's theorem provides a way of playing with the drift term of stochastic differential equations by changing the probability measure. We will use the theorem in the following situation.  Consider a filtration $\cF_t$ and probability measures $P$ and $Q$ on $\cF_\infty$ such that the restrictions are absolutely continuous $Q_t \ll P_t$.  Furthermore, let $P_t$ and $Q_t$ be such that the Radon-Nykodym derivative $D_t$ be continuous and of the form,
\begin{align}
  D_t = \frac{dQ_t}{dP_t} = \exp\left\{ L_t - \tfrac 12 \langle L,L \rangle_t \right\} \; 
 , 
\end{align}
for some local martingale $L_t$. 
If $M_t$ is a continuous $(\cF_t,P)$-local martingale then,
\begin{align}
  \widetilde{M}_t = M_t - \langle M, L \rangle_t \; ,
\end{align}
is a continuous $(\cF_t, Q)$-local martingale.  We refer the reader to
chapter VIII of \cite{book:Yor} for more details.
Note that in our application a local martingale $D_t \ge 0$, can be a
Radon-Nykodym 
derivative for some change of measure if and only if $\mE[ D_t ] = 1$,
i.e. it is a true martingale (Proposition VIII.1.13 of \cite{book:Yor}).

\section{Absolute Continuity}
\label{sec:abs}

In this section we argue that multiple SLEs which are absolutely
continuous wrt a single SLE (up to some stopping time) have the
correct quadratic variations for the BBK ansatz.  We contend this
is the correct notion for being ``locally like a single SLE''.  
\label{sec:singletomultiple}

\begin{figure}
\refstepcounter{figure}
\label{fig:H}
\begin{center}
\begin{tabular}{c}
\setlength{\unitlength}{1cm}
\begin{picture}(12,5)
  \put(3.85,3){
     \epsfxsize 5\unitlength
     \epsfbox{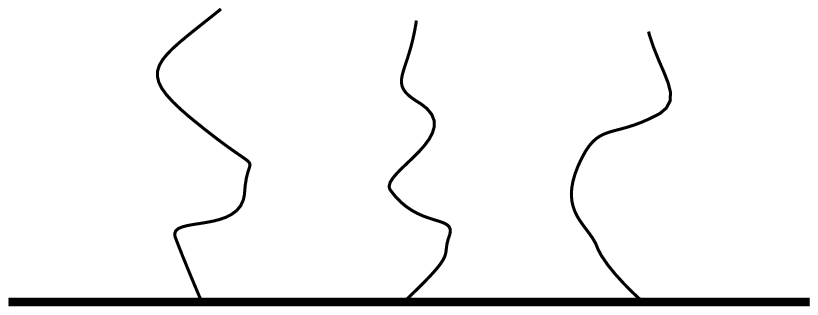}}
  \put(0,0){
     \epsfxsize 5\unitlength
     \epsfbox{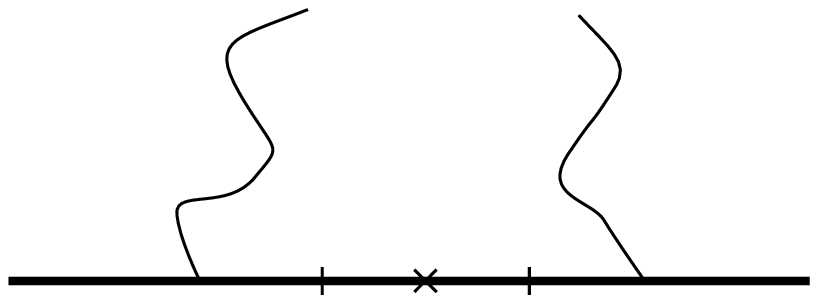}}
  \put(7,0){
     \epsfxsize 5\unitlength
     \epsfbox{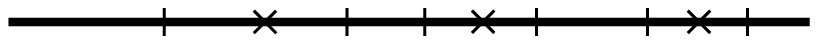}}
  \put(6.18,4.95){
     \epsfxsize 0.3\unitlength
     \epsfbox{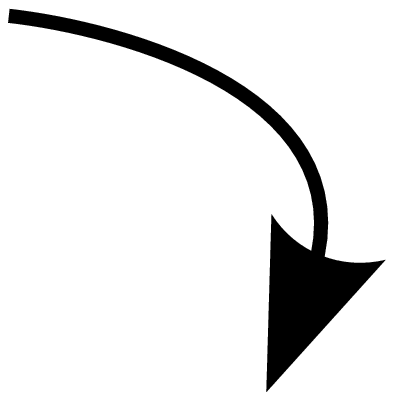}}
  \put(5.5,0.6){
     \epsfxsize 1\unitlength
     \epsfbox{figures/arrow_s_left.eps}}
  \put(4.2,1.75){
     \epsfysize 1\unitlength
     \epsfbox{figures/arrow_s_downright.eps}}
  \put(7.75,1.75){
     \epsfysize 1\unitlength
     \epsfbox{figures/arrow_s_downleft.eps}}
  \put(5.85,5.3){ \makebox(0,0)[l]{$\gamma^i$}}
  \put(4.85,2.1){ \makebox(0,0)[l]{$g_t^i$}}
  \put(7.65,2.1){ \makebox(0,0)[l]{$G_t$}}
  \put(5.8,1){ \makebox(0,0)[l]{$H_t^i$}}
\end{picture}
\vspace{0.5cm} \\
\parbox{12cm}{
\small\raggedright
{\em \begin{center} Figure \ref{fig:H} : A diagram representing the map $H_t^i$.\end{center}}}
\end{tabular}
\end{center}
\end{figure}

We have seen one natural way to encode multiple SLEs is in terms the single Loewner map $G_t$.  There is second, useful for situations involving time changes, which uses  a different Loewner map for each component curve.  In more detail, consider $n$-curves $\gamma_t^i$ each creating its own hull $K_t^i$ and let these hulls be rectified by maps $g_t^i : \mH / K_t^i \to \mH$ which define Loewner evolutions,
\begin{align}
  \dot{g}_t^i(z) = \frac{2 c_t^i}{g_t^i - w_t^i} \; , 
\qquad 
  g_0^i(z) = z \; . 
\qquad
  c_t^i = \frac{d }{dt}\hcap[ K_t^i ] \; .
  \label{eq:defn:single}
\end{align}
If all these curves are independent SLEs then the driving functions $w_t^i$ are independent martingales such that,
\begin{align}
  d\langle w^i,w^j \rangle_t = \kappa_i \, c_t^i \, \delta_{ij} \, dt \; .
\end{align}
To relate these driving functions to those of $G_t$, it is useful to define the maps $H_t^i$ (see figure \ref{fig:H}),
\begin{align}
  G_t = H_t^i \circ g_t^i \; . \label{eq:Hdefn}
\end{align}
It follows from this definition that,
\begin{align}
  x_t^i = H_t^i(w_t^i) \; ,
  \qquad
  a_t^i = {H_t^i}'(w_t^i)^2 c_t^i \; . \label{eq:afromc}
\end{align}
The first of these relations is trivial.  The second requires a little
more work which we reproduce from \cite{LawSchWer0209}.  Consider a
small increase in the length of the $i$th curve while keeping the
others fixed.  We need to compare the $\hcap$ of a small piece of
curve $\delta \gamma_t^i = g_t^i( \gamma_{t+\delta t}^i )$, $\hcap[
\delta \gamma_t^i ] \sim c_t^i \delta t$, with that of the image of this
curve under $H_t^i$.  Assuming the curve $\gamma_{t+\delta t}^i$ does
not intersect any of the other curves (near the time $t$), the map
$H_t^i$ is analytic near $\delta \gamma_t^i$ and we can approximate
$\hat{\delta \gamma}_t^i = H_t^i(\delta \gamma_t^i) \sim H_t^i(w_t^i)
+ {H_t^i}'(w_t^i) (\delta \gamma_t^i - w_t^i)$.  The result then
follows from the following easy properties of $\hcap$:  Let $g_A :
\mH / A \to \mH$ for some suitable $A \subset \mH$, $\lambda \ge 0$,
\begin{align} 
  \hcap[ A ] = \lim_{z \to \infty }\tfrac 12 z \, ( g_A( z ) - z) \; , 
\qquad
  \hcap[ \lambda A ] = \lambda^2 \hcap[ A ] \; .
\end{align}
Taking the time derivative of \eqref{eq:Hdefn} we obtain,
\begin{gather}
  \dot{H_t}(z) = \sum_j \frac{2 a_t^j}{H_t^i(z) - x_t^j} 
   - {H_t^i}'(z) \frac{2 c_t^i}{z - w_t^i} \; , \label{eq:Hdot}
\\
   \dot{H_t}(w_t^i) = -3 {H_t^i}''(w_t^i) c_t^i + \sum_{k \ne i} \frac{2 a_t^k}{x_t^i - x_t^k} 
\end{gather}
and hence using It\^o's formula,
\begin{align}
  dx_t = {H_t^i}'(w_t^i) dw_t^i
   + \left( \tfrac{\kappa_i}{2} -3 \right) {H_t^i}''(w_t^i) c_t^i dt 
   + \sum_{k \ne i} \frac{2 a_t^k}{x_t^i - x_t^k} dt
\end{align}
In particular the quadratic variations are,
\begin{align}
  d\langle x^i,x^j \rangle_t = {H_t^i}'(w_t^i)^2 \kappa_i c_t^i \delta_{ij} dt = \kappa_i a_t^i \delta_{ij} dt \; . \label{eq:quad}
\end{align}

Now, it is well known that under absolutely continuous changes of
measure the quadratic variation of a process will not change (for
example \cite{book:Yor}).  Hence for any (possibly stopped) multiple SLE
process absolutely continuous with respect to $n$-independent SLEs,
the quadratic variation of the driving functions will be given by
\eqref{eq:quad}.

\section{Conformal Invariance}
\label{sec:conformal}

We assume the driving functions satisfy a stochastic differential equation of the form, 
\begin{align}
  dx^i_t = dM_t^i + \sum_j Q_t^{ij}[x_t] a_t^j dt \; ,
\end{align}
where $M_t^i$ is a martingale with quadratic variation $d\langle M^i, M^i \rangle_t = \kappa_i a_t^i dt$.

\begin{figure}
\refstepcounter{figure}
\label{fig:Phi}
\begin{center}
\begin{tabular}{c}
\setlength{\unitlength}{1cm}
\begin{picture}(12,5)
  \put(0,3){
     \epsfxsize 5\unitlength
     \epsfbox{figures/3curves_1.eps}}
  \put(7,3){
     \epsfxsize 5\unitlength
     \epsfbox{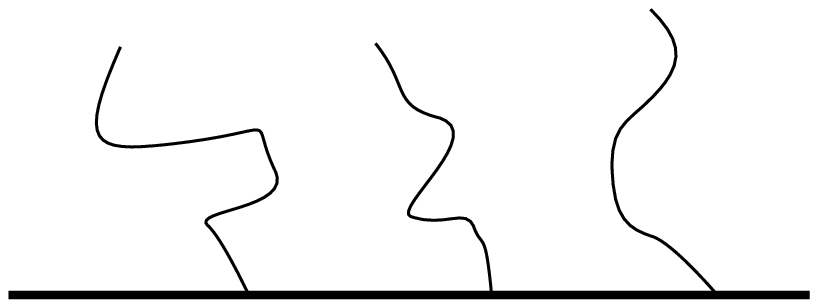}}
  \put(0,0){
     \epsfxsize 5\unitlength
     \epsfbox{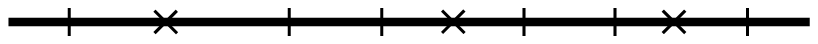}}
  \put(7,0){
     \epsfxsize 5\unitlength
     \epsfbox{figures/3curves_2a.eps}}
  \put(5.5,3.6){
     \epsfxsize 1\unitlength
     \epsfbox{figures/arrow_s_left.eps}}
  \put(5.5,0.6){
     \epsfxsize 1\unitlength
     \epsfbox{figures/arrow_s_left.eps}}
  \put(2.75,1.75){
     \epsfysize 1\unitlength
     \epsfbox{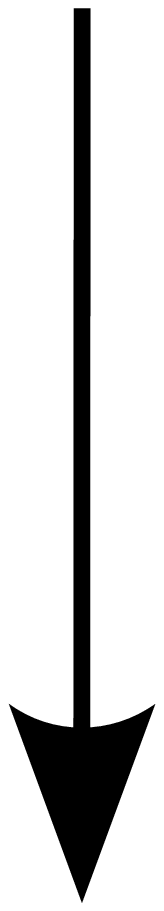}}
  \put(9.75,1.75){
     \epsfysize 1\unitlength
     \epsfbox{figures/arrow_s_down.eps}}
  \put(5.9,4){ \makebox(0,0)[l]{$h$}}
  \put(3,2.25){ \makebox(0,0)[l]{$G_t$}}
  \put(10,2.25){ \makebox(0,0)[l]{$\hat{G}_t$}}
  \put(5.8,1){ \makebox(0,0)[l]{$\Phi_t$}}
\end{picture}
\vspace{0.5cm} \\
\parbox{12cm}{
\small\raggedright
{\em \begin{center} Figure \ref{fig:Phi} : A diagram representing the map $\Phi_t$.\end{center}}}
\end{tabular}
\end{center}
\end{figure}

To see the effect of a conformal transformation on the form of these
equations we follow the arguments of \cite{LawSchWer0209}, generalised
to our situation.  Consider a M\"obius transformation $h$ of the UHP to
itself.  This map takes the curves $\gamma_t^i$ to new curves
$\hat{\gamma}_t^i = h( \gamma_t^i)$, which may be encoded by a Loewner
evolution with a new set of driving functions\footnote{We avoid the
  pathological case $h(\gamma_0^i) = \infty$.}, 
\begin{align}
  \dot{\hat{G}}_t  = \sum_i \frac{2 \hat{a}_t^i}{\hat{G}_t - \hat{x}_t^i} \; , 
\qquad 
  \hat{G}_0(z) = z \; .
\end{align}
Let $T = \inf\{ t : h(\infty) \in \hat{K}_t \}$ where $\hat{K}_t$ is the hull generated by $\cup_i \hat{\gamma}_t^i$.  For times $t<T$, the relation between the new driving functions and the old is encoded in the M\"obius map $\Phi_t$ (see figure \ref{fig:Phi}),
\begin{gather}
  \Phi_t := \hat{G}_t \circ h \circ G_t^{-1}  \; , \label{eq:def:phi}
\\
  \hat{x}_t^i = \Phi_t( x_t^i ) \; , 
\qquad
  \hat{a}_t^i = \Phi_t'( x_t^i )^2 a_t^i \; . \label{eq:xhat}
\end{gather} 
To obtain the second line we use an identical argument to that for equation \eqref{eq:afromc}.  It follows from \eqref{eq:def:phi} that,
\begin{align}
  \dot{\Phi}_t(z) = \sum_i \left[ \frac{ 2 \hat{a}_t^i }{\Phi_t(z) - \hat{x}_t^i} - \Phi'_t( z ) \frac{2 a_t^i}{z-x_t^i} \right] \; ,
\end{align}
and hence that,
\begin{align}
  \dot{\Phi}_t(x_t^i) = - 3 \Phi''_t( x_t^i ) a_t^i +  \sum_{k \ne i} \left[ \frac{ 2 \hat{a}_t^k }{\hat{x}_t^i - \hat{x}_t^k} - \Phi'_t( x_t^i ) \frac{2 a_t^k}{x_t^i-x_t^k} \right] \; .
\end{align}
Using these formulae we apply It\^o's formula to $\hat{x}_t^i$,
\begin{align}
  d\hat{x}_t^i 
  &{=}  \Phi_t'(x_t^i) dM^i_t {+} \!
  \left[
    \sum_{k \neq i} \frac{2 \hat{a}_t^k }{ \hat{x}_t^i -\hat{x}_t^k } 
    {+} \! \left( \tfrac{\kappa_i}{2}{-}3 \right)  \Phi_t''(x_t^i)  a_t^i
    {+} \Phi_t'(x_t^i)
     \! \left( \sum_j Q_t^{ij}[ x_t ] a_t^j {-} \sum_{k \neq i} \frac{2 a_t^k}{x_t^i -x_t^k} \right) 
  \right] \! dt \; .
\end{align}
For conformal invariance this new stochastic differential equation should have the same form as the original,
\begin{align}
  d\hat{x}_t^i = d\hat{M}_t^i + \sum_j Q_t^{ij}[ \hat{x}_t ] \hat{a}_t^j dt \; .
\end{align}
with the same functions $Q^{ij}_t$.  Equating the martingale parts we see,
\begin{align}
  d\hat{M}_t^i = \Phi_t'(x_t^i) dM^i_t \; ,
\end{align}
and hence $d\langle \hat{M}^i, \hat{M}^i \rangle_t = \kappa_i
\Phi_t'(x_t^i)^2 a_t^i dt = \kappa_i \hat{a}_t^i dt$ as required.
Equating the terms of finite variation we find,
\begin{align}
  \sum_j Q_t^{ij}[ \hat{x}_t ] \hat{a}_t^j -  \sum_{k \neq i} \frac{2 \hat{a}_t^k }{ \hat{x}_t^i -
  \hat{x}_t^k } &=  \left( \tfrac{\kappa_i}{2}{-}3 \right) \Phi_t''(x_t^i) a_t^i
  + \Phi_t'(x_t^i) \left( \sum_j Q_t^{ij}[ x_t ] a_t^j - \sum_{k \neq i} \frac{2 a_t^k}{x_t^i -
  x_t^k} \right) \; .
\end{align}
Writing
\begin{align}
  P^{ii}_t[x_t]  = Q_t^{ii}[ x_t ] \; , 
\qquad
  P^{ij}_t[x_t]  = Q_t^{ij}[ x_t ] - \frac{2}{x_t^i - x_t^j} \; , 
\end{align}
we find the objects $P^{ij}_t[x_t]$ transform under M\"obius transformations as,
\begin{align}
  P_t^{ii}[ \Phi_t(x_t) ] 
  = \left( \tfrac{\kappa_i}{2}{-}3 \right) \Phi_t''(x_t^i)
  + \Phi_t'(x_t^i) P_t^{ii}[ x_t ] \; ,
\qquad
  P_t^{ij}[ \Phi_t(x_t) ] = \Phi_t'(x_t^i) P_t^{ij}[ x_t ] \; .
  \label{eq:Ptrans}
\end{align}
and that conformal invariance requires the driving functions satisfy,
\begin{align}
  dx_t^i =  \sqrt{\kappa_i} dM_t^i+ \sum_j P_t^{ij}[x_t] a_t^j dt +  \sum_{k \neq i} \frac{2 a_t^k}{x_t^i -
  x_t^k} dt \; .
\end{align}
In the next section we find reparameterisation invariance places further constrains on $P_t^{ij}$. \\

So far in this section we have considered the functions
$Q_t^{ij}[x_t]$ depending only on the driving positions $x_t^i$.  In
more general situations, for example $SLE(\underline{\kappa},\underline{\rho})$
\cite{LawSchWer0209,Dub0303,Kyt0504}, one may wish to let $Q$ depend
on some extra parameters evolving via $G_t$, $Q_t^{ij}[ x_t , y_t ]$,
where $y_t^\ell = G_t( y_0^\ell )$, $\ell = 1, \ldots, r$.  One can
check that this does not affect our result as $\hat{y}_t^\ell =
\Phi_t( y_t^\ell )$ and the new $P_t^{ij}$ again satisfying, 
\begin{align}
    P_t^{ii}[ \Phi_t(x_t) , \Phi_t(y_t) ] 
  &= \left( \tfrac{\kappa_i}{2}{-}3 \right) \Phi_t''(x_t^i)
  + \Phi_t'(x_t^i) P_t^{ii}[ x_t, y_t ] \; ,
\\
  P_t^{ij}[ \Phi_t(x_t), \Phi_t(y_t) ] &= \Phi_t'(x_t^i) P_t^{ij}[ x_t, y_t ] \; .
\end{align}
It is important to note that in our definition of conformal invariance we have
implicitly assumed none of the curves go to infinity.  In recovering
this eventuality we obtain a particular example of an
$SLE(\kappa,\rho)$ process with $r=1$, $y_t = G_t(\infty) = \infty$
and $\Phi_t(y_t) = \hat{G}_t(h(\infty))$.
Another generalisation involving Lie groups following
\cite{BetGruLudWie0503} will be considered in future work.

\section{Reparameterisation Invariance}
\label{sec:commute}

The consequences of reparameterisation invariance for multiple SLEs
was first considered by Dub\'edat in \cite{Dub0411,Dub0507} wherein a set of
equations that are necessary for the driving function of an SLE to be
reparameterisation invariant where derived.  For completeness, we will
rederive these equations and apply them to our conformally invariant
evolution.  In so doing we will obtain some interesting results
previously assumed from conformal field theory. \\ 

To begin, we consider the driving function,
\begin{align}
  dx_t^i = dM_t^i + \sum_{j} Q_{ij}[x_t] a_t^j dt
\end{align}
Note that such a process implicitly includes the $SLE(\kappa,\rho)$
process (with $r{=}1$, $y_t {=} G_t(\infty) {=} \infty$) that describes
curves going to infinity.
As a Markov process it has the infinitesimal generator,
\begin{align}
   \cL_t = \sum_i a_t^i \left[ \frac{\kappa^i}{2}
    \frac{\partial^2}{{\partial x_i}^2 }
 + \sum_j Q_{ji}[x]
  \frac{\partial}{\partial x_j }
 \right]
  = \sum_i a_t^i \cD_i 
\end{align}

To find Dub\'edat's condition consider two of the multiple SLE curves,
$i$ and $j$ say, and chose the time parameterisation such that one
first grows the curve $i$ until $\hcap[ K_t^i
] = \eps_i$,  one then grows the curve $j$ until $\hcap[ K_t^j ] =
\eps_j$.  An equally valid time parameterisation of the resulting
system can be obtained by growing $j$ and then $i$.  Dub\'edat's
condition arises from equating the expectations of observables
obtained by both ways of growing $i$ and $j$.

Let us begin by growing curve $i$ and then $j$ such that both $\eps_i$ and $\eps_j$ are infinitesimal (they could be equal).  Let $\delta$ be the time at which the curve $i$ has grown to size $\eps_i$ and $2\delta$ be when both curves have finished growing.  
We will need (using \eqref{eq:afromc} and \eqref{eq:defn:single}),
\begin{align}
\begin{array}{ll}
  a_0^i\delta = c_0^i\delta  = \eps_i \; , & a_0^k = 0 \; , k \ne i \\
  a_\delta^j \delta = {H_\delta^j}'(w_\delta^j)^2 c_\delta^j \delta =  {H_\delta^j}'(w_t^j)^2 \eps_j \; , & a_\delta^k = 0 \; , k \ne j 
\end{array}
\end{align}
where in this case, the map $H_\delta^j$ is the Loewner map of the curve $i$ at time $\delta$ and so is given by,
\begin{gather}
  H_\delta^j(z) = z + \frac{ 2 \eps_i }{ z - x_0^i } + \cO(\eps_i^2) \; , 
\qquad
  x_0^j = x_\delta^j + \cO(\eps_i) = w_\delta^j + \cO(\eps_i)
\\
  {H_\delta^j}'(w_\delta^j)^2 = 1 - \frac{ 4 \eps_i }{ (x_0^j - x_0^i)^2 } + \cO(\eps_i^2) 
\end{gather}
Now consider the expectation of some functional at time $2\delta$.  Using the infinitesimal generator this can be expanded to second order in $\eps_i$ and $\eps_j$,
\begin{align}
 \mE[ & f( \gamma_{2\delta} ) | x_0 ] = 
  \mE[ \mE[ f( \gamma_{2\delta} ) | x_\delta ] | x_0 ]=
  \mE[ ( 1 + \delta \cL_\delta + \tfrac{\delta^2}{2} \cL_\delta^2) f( \gamma_{\delta}  ) | x_0 ] \notag \\
   &= ( 1 + \delta \cL_0 + \tfrac{\delta^2}{2} \cL_0^2) ( 1 + \delta \cL_\delta + \tfrac{\delta^2}{2} \cL_\delta^2) \mE[ f( \gamma_0 ) | x_0 ] \notag \\
  &= 
  \left[ 1 + \delta a_0^i \cD_i  + \frac{\delta^2}{2} {a_0^i}^2 \cD_i^2 + \ldots 
\right] 
  \left[ 1 + \delta a_\delta^j \cD_j + \frac{\delta^2}{2} {a_\delta^j}^2 \cD_j^2 + \ldots \right] \mE[ f( \gamma_0 ) | x_0 ] \notag \\
  &= \left[ 
  1 + \eps_i \cD_i {+} \eps_j \cD_j - \frac{4 \eps_i \eps_j}{(x_i - x_j)^2} \cD_j {+} \frac{\eps_i^2}{2} \cD_i^2 {+} \eps_i \eps_j \cD_i \cD_j {+} \frac{\eps_j^2}{2} \cD_j^2 + \ldots \right]  \mE[ f( \gamma_0 )| x_0  ] \; .
\end{align}
Equating this expression with that obtained by growing $j$ and then $i$ we obtain Dub\'edat's commutation relations,
 \begin{align}
   [ \cD_i , \cD_j ] =  \frac{ 4 }{(x_i-x_j)^2} ( \cD_j - \cD_i )
\end{align}
or in components,
\begin{align}
\sum_n \left[ \kappa^i \frac{ \partial Q_{nj} }{\partial x_i } 
  \frac{\partial}{\partial x_i} 
  - \kappa^j \frac{ \partial Q_{ni} }{\partial x_j } 
  \frac{\partial}{\partial x_j} 
+ \frac{ \kappa^i}{2} \frac{\partial^2 Q_{jn}}{
      {\partial x_i}^2 } 
     - \frac{ \kappa^j}{2} \frac{\partial^2 Q_{ni}}{ {\partial x_j}^2
     } 
  + \sum_\ell \left( Q_{\ell i} \frac{\partial Q_{nj}}{\partial x_\ell}
  -  Q_{\ell j} \frac{\partial Q_{ni}}{\partial x_\ell} \right)
  \right] \frac{\partial}{\partial x_n}
\notag \\
 -\frac{4}{(x_i-x_j)^2} \left[
  \frac{\kappa^j}{2} \frac{\partial^2}{{\partial x_j}^2 }
  -  \frac{\kappa^i}{2} \frac{\partial^2}{{\partial x_i}^2 }
  + \sum_n \left( Q_{nj} - Q_{ni} \right) \frac{\partial}{\partial
    x_n} \right] = 0 \hspace{10mm}
  \label{eq:Dincomp}
\end{align}
Applying these constraints to our conformal evolution, we first consider the constraints arising from the second order terms. Reintroducing $P^{ij}$ (and assuming $\kappa^i \ne 0$) we find,
\begin{align} 
   \frac{\partial P^{nj}[x]}{\partial x_i} = 0 \, \text{ for }
   n \ne j  \, , 
\qquad
  \kappa^i \frac{\partial P^{jj}[x]}{\partial x_i}
  = \kappa^j \frac{\partial P^{ii}[x]}{\partial x_j} \; ,
\end{align}
The first of these relations imply that $P^{ij}[x]$ is a function of $x_j$ only.  However, this is only consistent with the conformal transformation \eqref{eq:Ptrans} if $P^{ij} = 0$.
The second equation is an integrability condition,
\begin{align}
  P^{ii} = \kappa^i \frac{\partial}{ \partial x_i} F[ x ] \; . \label{eq:biians}
\end{align}
Moving to first order we first notice that terms with $n \ne i \text{ or }
j$ are now trivially satisfied.
In the case $n = i$ we find after a little algebra,
\begin{align}
  0 &= \kappa^i \frac{\partial}{\partial x_i} \left[
   - \frac{1}{(x_j - x_i)^2}
   - \frac{\kappa^j}{2} \left[ 
        \frac{\partial^2 F[x]}{\partial x_j^2} 
        {+}\! \left[ \frac{\partial F[x]}{\partial x_j} \right]^2 \right]
   + \sum_{\ell \ne j} \frac{2}{x_j - x_\ell} \frac{\partial F[x]}{\partial x_\ell} 
   + \frac{1}{\kappa^i} \frac{6}{(x_i - x_j)^2} 
  \right]
\end{align}
Writing $F[x] = \log Z[x]$,  $h_i = \tfrac{6 - \kappa_i}{2\kappa_i}$ and letting $i$ range over $i \ne j$ we see,
\begin{align}
    A_j( x_j ) = 
   \sum_{\ell \ne j} \frac{2 h_\ell}{(x_j - x_\ell)^2}
   - \frac{\kappa^j}{2} \frac{1}{Z[x]} 
        \frac{\partial^2 Z[x]}{\partial x_j^2} 
   + \sum_{\ell \ne j} \frac{2}{x_j - x_\ell} \frac{1}{Z[x]}\frac{\partial Z[x]}{\partial x_\ell} 
\end{align}
However, the conformal properties of $P^{ii}$ imply that under M\"obius transformations,
\begin{align}
  Z[ \Phi(x) ] = Z[ x ] \prod_i \Phi'(x_i)^{-h_i} \; , \label{eq:Ztransform}
\end{align}
and hence scale and translational covariance (for example) require $A_j(x_j)=0$. \\

In conclusion, we see that conformal invariance together with reparameterisation invariance imply that the driving function for multiple SLEs should have the form,
\begin{align}
  dx_t^i = dM_t^i 
   + \kappa_i \frac{\partial}{\partial x_t^i} \log Z[x_t] dt
   + \sum_{k \ne i} \frac{2 a_t^k}{x_t^i - x_t^k} dt \; , \label{eq:x}
\end{align}
where the function $Z[x]$ transforms as \eqref{eq:Ztransform} under M\"obius transformations and satisfies the following so called null vector equations,
\begin{align}
   0 = \frac{\kappa_i}{2} \frac{\partial^2 Z[x]}{\partial x_i^2} 
   - \sum_{k \ne i} \frac{2}{x_i - x_k} \frac{\partial Z[x]}{\partial x_k} 
   - \sum_{k \ne i} \frac{2 h_k}{(x_i - x_k)^2} Z[x] \; , \label{eq:nve}
\end{align}
This is precisely the proposal made by Bauer, Bernard and Kyt\"ol\"a \cite{BauBerKyt0503}. \\

We end this section by proving the following simple theorem,\\

\noindent {\bf Proposition:} {\em There exists a $\kappa$ such that the only non-trivial solutions to \eqref{eq:nve} have $\kappa_i \in \{ \kappa, \tfrac{16}{\kappa} \}$ for all $i$.} \\

\noindent To prove it, let us write the null vector equation using a differential operator $\cO_i Z[x] = 0$ then,
\begin{align}
  [\cO_i, \cO_j]-\frac{4}{(x_i - x_j)^2} \left( \cO_j - \cO_i \right)
  = -\frac{3 ( \kappa_i - \kappa_j )(16 - \kappa_i \kappa_j )}{\kappa_i \kappa_j (x_i - x_j)^4}
\end{align}
and $Z[x]$ can only be a simultaneous solution to equations $i$ and $j$ if the proposition holds.\\

As observed in previous work on SLEs, it is natural to associate a
$\kappa$ SLE with a $\phi_{1,2}$ field in conformal field
theory\footnote{See \cite{BauBer0210,BauBerKyt0503} for more details
  on notation and motivation.}.  In this case an SLE with
$\tfrac{16}{\kappa}$ is 
naturally associated with a $\phi_{2,1}$ field.  We see here that it
is quite consistent to put both processes together in the same
geometry.  Furthermore, by looking at the singularities of the
solutions to the null vector equations one can study the fusion of
conformal operators.  In particular, we now have all the building
blocks to construct a probability interpretation for the full Kacs
table $\phi_{r,s}$.  We leave more detailed discussion of this to
future work. 

Also note that what we have shown is that it is not consistent to put
two SLEs together unless their $\kappa$ parameters are related by the
proposition.  One could say this is the probability theory realisation
of an observation from physics that it is not possible to build a
conformal field theory using representations from Virasoro algebras
with different central charges,
$c=\tfrac{1}{2\kappa}(6-\kappa)(3\kappa -8)$.

\section{Time Changes Take Two}
\label{sec:timestwo}

So far in this paper we have found it most natural to use the driving
functions $x_t^i$ to describe our multiple SLE processes.  When
considering time changes however, it is sometimes better to use the
$w_t^i$ as defined in section \ref{sec:singletomultiple}. 
In this section we will study conformal multiple SLEs
by taking n-independent SLEs with driving functions $w_t$ and
conditioning them to satisfy \eqref{eq:x}.  We will do this in two
steps: first we will condition the $n$-independent SLEs to move in
the background of the other SLEs, then we introduce the drift term
involving $Z[x]$.  
As a consequence of this procedure, we will be able to study directly
the consequences of reparameterisation invariance. \\

\begin{figure}
\refstepcounter{figure}
\label{fig:phitt}
\begin{center}
\begin{tabular}{c}
\setlength{\unitlength}{1cm}
\begin{picture}(12,5.5)
  \put(0,3){
     \epsfxsize 5\unitlength
     \epsfbox{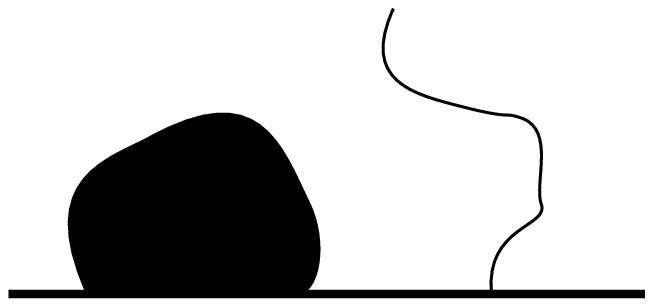}}
  \put(7,3){
     \epsfxsize 5\unitlength
     \epsfbox{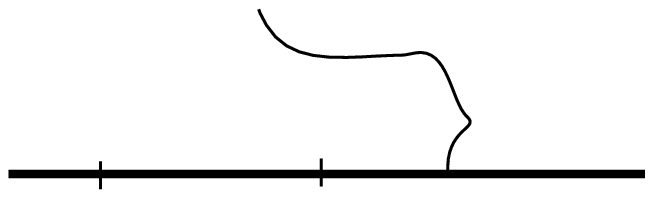}}
  \put(0,0){
     \epsfxsize 5\unitlength
     \epsfbox{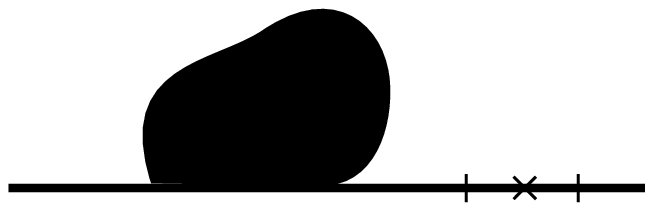}}
  \put(7,0){
     \epsfxsize 5\unitlength
     \epsfbox{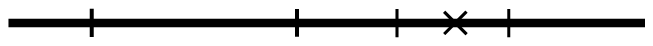}}
  \put(5.5,3.6){
     \epsfxsize 1\unitlength
     \epsfbox{figures/arrow_s_left.eps}}
  \put(5.5,0.6){
     \epsfxsize 1\unitlength
     \epsfbox{figures/arrow_s_left.eps}}
  \put(2.75,1.75){
     \epsfysize 1\unitlength
     \epsfbox{figures/arrow_s_down.eps}}
  \put(9.75,1.75){
     \epsfysize 1\unitlength
     \epsfbox{figures/arrow_s_down.eps}}
  \put(1.5,4.65){
     \epsfysize 0.3\unitlength
     \epsfbox{figures/arrow_down.eps}}
  \put(1.15,1.45){
     \epsfysize 0.2\unitlength
     \epsfbox{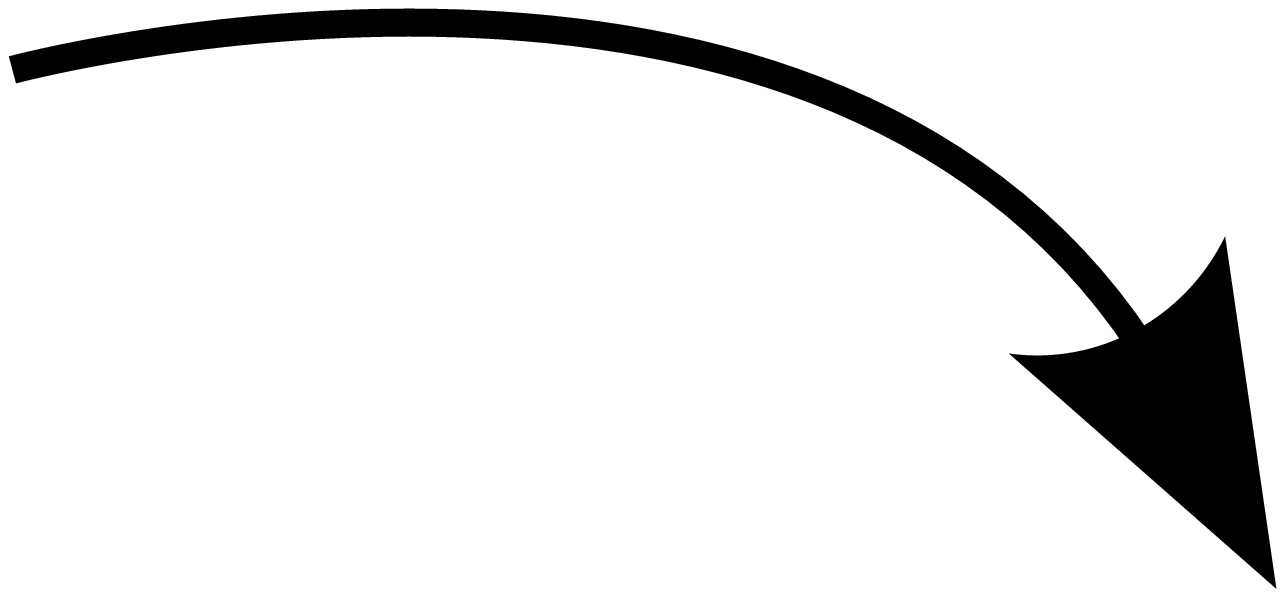}}     
  \put(5.9,4){ \makebox(0,0)[l]{$h$}}
  \put(3,2.25){ \makebox(0,0)[l]{$g_t$}}
  \put(10,2.25){ \makebox(0,0)[l]{$\hat{g}_t$}}
  \put(5.8,1){ \makebox(0,0)[l]{$\phi_t$}}
  \put(1.15,4.85){ \makebox(0,0)[l]{$A$}}
  \put(0.2,1.6){ \makebox(0,0)[l]{$g_t(A)$}}
\end{picture}
\vspace{0.5cm} \\
\parbox{12cm}{
\small\raggedright
{\em \begin{center} Figure \ref{fig:phitt} : A diagram representing the map $\phi_t$.\end{center}}}
\end{tabular}
\end{center}
\end{figure}

Before we start with the multiple SLEs however, it is helpful to take
a moment to reconsider a single SLE and it's image under a conformal
map. 
Consider some hull $A \subset \mH$ away from $w_0$ with map $h: \mH /
A \to \mH$. 
Let $\gamma_t$, $g_t$, $w_t$ and $c_t$ be a standard SLE in the UHP as
defined in section \ref{sec:singletomultiple} and define the stopping
time $T = \inf\{ t \, : \, \gamma_t \in A \}$.  
Let $\hat{\gamma_t} = h(\gamma_t)$, $\hat{g}_t$, $\hat{w}_t$ and
$\hat{c}_t$ be the image of the SLE under the map $h$.  As in section
\ref{sec:conformal} we use (see figure \ref{fig:phitt}), 
\begin{gather}
  \phi_t = \hat{g}_t \circ h \circ g_t^{-1} \; ,
\qquad
  \hat{w}_t = \phi_t(w_t) \; ,
\qquad
  \hat{c_t} =  \phi_t'(w_t)^2 c_t \; ,
\\
  \dot{\phi}_t(w_t) = -3 \phi_t''(w_t) \; ,
\qquad
  \dot{\phi}_t'(w_t) = \frac{\phi_t''(w_t)^2}{2\phi_t'(w_t)}-\frac{4\phi_t'''(w_t)}{3} \; ,
 \label{eq:phiformulae}
\end{gather}
to find that the image driving function satisfies,
\begin{align}
  d\hat{w}_t = \phi_t'(w_t) dw_t 
  + \left( \tfrac{\kappa}{2} - 3 \right)\phi_t''(w_t) c_t dt  \; .
\end{align}

Now we would like to change the measure such that this new process is an SLE.  This is achieved by applying Girsanov's theorem: if we denote the old measure by $P$, the required new measure is $P_\text{new} = P D_t$ with Radon-Nykodym derivative\footnote{It is easy to check that $D_t$ is a local martingale, however, to apply Girsanov's theorem we must show that $D_t$ is a true martingale.  We will discuss this at the end of the section.},
\begin{align}
  D_t &= \exp \left\{ h \int_0^t \frac{\phi_s''(w_s)}{\phi_s'(w_s)} dw_s 
          - \frac{h^2 \kappa}{2} \int_0^t \frac{\phi_s''(w_s)^2}{\phi_s'(w_s)^2} c_s ds  \right\} 
\notag \\
     &= \frac{ \phi_t'(w_t)^h }{ h'(w_0)^h } 
        \exp \left\{ -\frac{c}{6} \int_0^t S\phi_s(w_s) c_s ds \right\} \; ,
  \label{eq:restrict}
\end{align}
wherein $S$ denotes the Schwarzian derivative,
\begin{align}
  S\phi(z) = \frac{ \phi'''(z) }{ \phi'(z) } - \frac 32 \frac{\phi''(z)^2}{\phi'(z)^2}\; ,
\qquad
  h = \frac{6-\kappa}{2\kappa}\; ,    
\qquad
  c = \frac{(6-\kappa)(3\kappa - 8)}{2 \kappa}   \label{eq:centralcharge} 
\end{align}
To obtain the second line in \eqref{eq:restrict}, apply It\^o's formula to $\log \phi_t'(w_t)$ with the help of the formulae in \eqref{eq:phiformulae}.  Note that \eqref{eq:restrict} is the restriction martingale of \cite{LawSchWer0209} and $c$ is the central charge.  

Under the new measure, $\hat{w}_t$ is a martingale and $\hat{\gamma}_t$ is a ordinary SLE in the UHP.  By conformal invariance of SLE, the image of this process under the inverse map $h^{-1}$ is an ordinary SLE in the non-standard domain $\mH / A$.  Hence the process $w_t$ under the measure $P D_t$ describes an SLE in the domain $\mH / A$.

We would like to use this to condition our $n$-independent SLEs such that each curve evolves as an SLE living in the background created by the other curves.  
More precisely, let ${K_t^i}^c$ be the hull created by all the curves except $\gamma_t^i$ and let $h_t^i$ be the map $h_t^i :  \mH / {K_t^i}^c \to \mH$.  As usual, let $g_t^i$ be the Loewner map for the curve $\gamma_t^i$  and let $\hat{g}_t^i$ be the Loewner map for the image of $\gamma_t^i$ under $h_t^i$.  Following the single curve example we define,
\begin{align}
  H_t^i = \hat{g}_t^i \circ h_t^i \circ {g_t^i}^{-1}
\end{align}
and note that under an infinitesimal increase in the time along the $i$th curve $t^i$, while keeping the others fixed, the driving function of the image satisfies,
\begin{align}
  d^i\hat{w}_t^i = {H_t^i}'(w_t^i) dw_t^i 
  + \left( \tfrac{\kappa}{2} - 3 \right) {H_t^i}''(w_t^i) c_t^i dt^i \; ,
\qquad
  \hat{c}_t^i = {H_t^i}'(w_t^i)^2 c_t^i \; . 
\end{align}
For this image process to be an SLE we need to change the measure to remove the drift term.  By Girsanov's theorem, the Radon-Nykodym derivative is given by,
\begin{align}
  D_t^i = 
  \exp\left\{ h_i 
  \int_0^t \frac{{H_s^i}''(w_s^i)}{{H_s^i}'(w_s^i)} dw_s^i
  - \frac{h_i^2 \kappa_i}{2} \int_0^t \frac{{H_s^i}''(w_s^i)^2}{{H_s^i}'(w_s^i)^2} c_s^i ds \right\}
\end{align}
and so the process $w_t^i$ with the measure $P D_t^i$ is an SLE moving in the background of the other curves.

Now we grow all the curves simultaneously.  As each curve grows we need to adjust the measure to keep it moving in the background of the others.  The new conditioned measure is,
\begin{align}
  P_\text{new} =  P \prod_i D_t^i = P \prod_i {H_t^i}'(w_t^i)^{h_i}
  \exp \left\{ - \frac{c_i}{6} \int_0^t SH_s^i(w_s^i) c_s^i ds
  + \sum_{ k \ne i } \int_0^t \frac{ 2 \, h_i a_s^k }{ (x_s^i - x_s^k)^2} ds
  \right\} \; . \label{eq:Pnew}
\end{align}
To obtain the RHS, we have used It\^o's formula applied to $\log {H_t^i}'(w_t^i)$, used equation \eqref{eq:Hdot} to calculate ${\dot{H_t^i}}'( w_t^i )$ and written $x_t^i = H_t^i(w_t^i)$.  There is no denominator because $H_0^i(z) = z$.
The effect of this new measure on the driving functions for the Loewner map $G_t$ is particularly striking.  From section \ref{sec:singletomultiple} we recall,
\begin{align}
  G_t = H_t^i \circ g_t^i \; ,
\end{align}
and the driving functions for the multiple SLE are related to those of the single SLEs by,
\begin{gather}
  dx_t^i = {H_t^i}'(w_t^i) dw_t^i
   + \left( \tfrac{\kappa_i}{2} -3 \right) {H_t^i}''(w_t^i) c_t^i dt 
   + \sum_{k \ne i} \frac{2 a_t^k}{x_t^i - x_t^k} dt \; . \label{eq:dx}
\end{gather}
This is under the measure $P$.  Under the new measure $P_\text{new}$ it is easy to check that,
\begin{align}
  N_t^i =  \int_0^t {H_s^i}'(w_s^i) dw_s^i
   + \left( \tfrac{\kappa_i}{2} -3 \right) \int_0^t {H_s^i}''(w_s^i) c_s^i ds 
\end{align}
is a martingale with quadratic variation $d\langle N^i, N^i \rangle_t = \kappa_i a_t^i dt$ and equation \eqref{eq:dx} may be rewritten,
\begin{align}
  dx_t^i = dN_t^i + \sum_{k \ne i} \frac{2 a_t^k}{x_t^i - x_t^k} dt \; .
   \label{eq:cardys}
\end{align}
This is precisely the system derived by Cardy in \cite{Car0310}.  The advantage of our derivation is we know how the new driving functions are related to those of the single SLEs and so it is easier to consider the effects of time changes. \\

The final step in building our conformally invariant process is adjust the measure to introduce the conformal drift term.  Using Girsanov's theorem again, this is achieved with the Radon-Nykodym derivative,
\begin{align}
    C_t &= \exp\left\{ \sum_i \int_0^t \frac{\partial}{\partial x_s^i} \log Z[x_s] dN_s^i 
  - \frac {1}2 \sum_i \int_0^t \left( \frac{\partial}{\partial x_s^i} \log Z[x_s] \right)^2 \kappa_i a_s^i ds  \right\} \\
    &= \frac{Z[x_t]}{Z[x_0]} \exp \left\{ -  \sum_i \int_0^t \frac{1}{Z[x_s]} \left[ 
    \sum_{k \ne i} \frac{2}{x_s^k - x_s^i} \frac{\partial Z[x_s]}{\partial x_s^k} + \frac{\kappa_i}{2} \frac{\partial^2 Z[x_s]}{{\partial x_s^i}^2}  \right] a_s^i ds \right\}
\end{align}
where as usual, we have used It\^o's formula on $\log Z[x_t]$. \\

To summarise, we have shown that under the measure $Q = P_\text{new} C_t = P C_t \prod_i D_t^i$, the processes $x_t^i$ satisfy,
\begin{align}
  dx_t^i = dM_t^i + \kappa_i \frac{\partial}{\partial x_t^i} \log Z[x_t] a_t^i dt
    + \sum_{k \ne i} \frac{2 a_t^k}{x_t^i - x_t^k} dt
\end{align}
were the $M_t^i$ are $Q$-martingales. \\

After all that, let us now return to the question of time changes.
Consider the expectation value of some functional of our curves with respect to the conformal multiple SLE measure $Q$,
\begin{align}
  \mE_Q [ f( \gamma_t ) ] = \mE_{P} [ C_t \prod_i D_t^i \, f( \gamma_t ) ] \; .
\end{align}
Since the original measure $P$ is reparameterisation invariant, the expectation of an (invariant) object will be invariant if $C_t \prod_i D_t^i$
is also invariant.  The first factor in each term is fine because it only depends on the endpoint, this leaves the contribution from the exponentials:
\begin{align}
  - \sum_i \! \frac{c_i}{6} \! \int_0^t \!\! SH_s^i(w_s^i) c_s^i ds
  + \sum_i \! \int_0^t \!\! \frac{1}{Z[x_s]} \! \left[ 
   \sum_{k \ne i} \frac{ 2 \, h_k  Z[x_s] }{ (x_s^k {-} x_s^i)^2}
   {-} \sum_{k \ne i} \frac{2}{x_s^k {-} x_s^i} \frac{\partial Z[x_s]}{\partial x_s^k} 
   {-} \frac{\kappa_i}{2} \frac{\partial^2 Z[x_s]}{{\partial x_s^i}^2}  \right] a_s^i ds 
\end{align}
Collecting terms in this way it is clear the second integral is invariant if the integrand vanishes.  This is the null vector equation for $Z[x]$.
Turning to the first term, we introduce coordinates on each curve $t^i(s) = \hcap[ K_s^i ]$ 
so that,
\begin{align}
  c_s^i = \frac{dt^i(s)}{ds} 
\end{align}
and the object of interest becomes the integral of a one-form,
\begin{align}
   dR = - \sum_i \frac{c_i}{6} SH_s^i(w_s^i) \, dt^i
\end{align} 
For invariance this integral should not depend on the integration path in ``time space''.  This will be true if and only if the form is closed.  Noting that,
\begin{gather}
  \frac{\partial H_s^i(w_s^i)}{\partial t^k} =  \frac{2 {H_s^k}'(w_s^k)^2 }{x_s^i - x_s^k} \; ,
\quad
  \frac{\partial {H_s^i}'(w_s^i) }{\partial t^k} =  \frac{2 {H_s^k}'(w_s^k)^2  {H_s^i}'(w_s^i)}{ (x_s^i - x_s^k)^2 }  \; , \quad \ldots \\
  \frac{\partial  }{\partial t^k}SH_s^i(w_s^i) = - \frac{12 {H_s^k}'(w_s^k)^2 {H_s^i}'(w_s^i)^2 }{ (x_s^i - x_s^k)^4 } \; ,
\end{gather}
the form is closed if and only if,
\begin{align}
  c_i = c_j \; , \text{ for all $i$ and $j$.}
\end{align}
Recalling the definition of the central charge $c_i$, \eqref{eq:centralcharge}, this is true if and only if,
\begin{align}
  \kappa_i = \kappa_j \; , \quad \text{ or } \quad 
  \kappa_i = \frac{16}{\kappa_j} \; . \label{eq:kappacond}
\end{align}
However, we saw in section \ref{sec:commute} that this is also a consequence of the null vector equation so the null vector is sufficient for reparameterisation invariance. \\

So what have we gained?  For one, the calculation we have done here is
more general than that of section \ref{sec:commute} in that it can be generalised to
cases where the object $Z$ does not define a Markov process.  This
could be useful in generalising SLE techniques to statistical models
with non-conformal boundary conditions. 

On the other hand, our method here is not yet a rigorous derivation
of necessary conditions for reparameterisation invariance, unlike
section \ref{sec:commute}.  To use Girsanov's theorem properly we need
Radon-Nykodym derivatives which are true martingales and not just
local.  Local martingales are true martingales up to some stopping
time, and so the above is valid for suitably stopped processes.   

Let us look at this directly.  Assuming the null vector equation, our
RN-derivative is, 
\begin{align}
  C_t \prod_i D_t^i = 
  \frac{Z[x_t]}{Z[x_0]} \prod_i {H_t^i}'(w_t^i)^{h_i}
  \exp \left\{ - \frac{c_i}{6} \int_0^t SH_s^i(w_s^i) c_s^i ds
  \right\} \; .
  \label{eq:RNder}
\end{align}
From \cite{LawSchWer0209}, we know $0 \le {H_t^i}'(w_t^i) \le 1$, $SH_t^i(w_t^i) \le 0$ and that
 ${H_t^i}'(w_t^i) \to 0$ 
as the curve $\gamma_t^i$ approaches ${K_t^i}^c$.  This means that our
RN-derivative is well defined and a true martingale so long as we stop
the process before the curves intersect.  Another way of saying this
is that the multiple SLE processes are absolutely continuous wrt
$n$-independent SLEs away from points where the curves collide or
intersect. However, it is precisely the points where the curves
collide or bounce off each other that we are most interested in. 

For $c_i \le 0$ and $h_i \ge 0$, the product term in \eqref{eq:RNder}
is bounded.  This requires $\kappa_i \le \tfrac 83$ or $\kappa_i = 6$.
For the remainder of this section we will concentrate on the range
$\kappa_i = \kappa \le \tfrac 83$.   We will also work with the
particular parameterisation $a_t^i = 1$.  This is useful since with this choice, the hull created by a single curve cannot enclose the tip of a second curve.  Furthermore, if two curves collide, then the hull created by their union will not contain
any other pairs of (growing) curves (We demonstrate this in the appendix).  Because of these observations, we order our driving functions $x_t^1 \le x_t^2 \le \ldots \le x_t^n$ and note that two curves collide at their tips if and only if $x_t^i \to x_t^{i+1}$.  Also note that $\hcap[K_t] = n t$ and hence a curve only reaches the point $\infty$ in infinite time.

From the application of conformal field theory, the solutions to the null vector equations are well understood \cite{BelPolZam84,DotFat84} and form a finite dimensional vector space.   BBK \cite{BauBerKyt0503} argued that among the vectors in this space are a set which can be identified with the possible topological configurations of curves.  
To see part of this picture, we recall that solutions (for $\kappa_i = \kappa \le \tfrac 83$) can have two possible singular behaviours as two points come together:
\begin{align}
  Z[x] &\sim C_3 (x_i - x_{i+1})^{\tfrac 2\kappa} Z_3[ x_1, \ldots, x_{i-1}, \tfrac 12(x_i + x_{i+1}), x_{i+2}, \ldots,x_n]\; , \label{eq:Zsing3}\\
  Z[x] &\sim C_1 (x_i - x_{i+1})^{-2h} Z_1[ x_1, \ldots, x_{i-1}, x_{i+2}, \ldots,x_n]\;. \label{eq:Zsing1} 
\end{align}
Here $C_1$ and $C_3$ are constants and the function $Z_3[x]$ is known to satisfy a third order null vector equation and is possibly singular when it's arguments come together or go to infinity.  The function $Z_1[ x_1, \ldots, x_{i-1}, x_{i+2}, \ldots,x_n]$ actually satisfies the same type of null vector equations as $Z$, but for $n{-}2$ variables.

We now return to the RN-derivative.
First we will show that with probability one, multiple SLE processes with $\kappa_i = \kappa \le \tfrac 83$ will only collide at their endpoints.  
Let $I_t^i = \inf_{s<t}\{ x_s^{i+1} -
x_s^i \}$ and let $I_t = \min_i\{ I_t^i \}$.  Now also recall that $T$ is the first time that two different curves meet and so it follows that in our chosen parameterisation, the event $\{ T<\infty$, $I_T = 0 \}$ signifies that (at least) two curves meet at their tips while for $\{ T<\infty$, $I_T > 0 \}$ two curves will meet away from their endpoints.  From \cite{LawSchWer0209} we know then that if the curve $i$ meets another curve at time $T < \infty$ then $\lim_{t\to T} {H_t^i}'(w_t^i) = 0$ and hence from the formula for the RN-derivative $\mE_Q[ \One_{\{ T<\infty \, , \; I_T > 0 \}} ] = 0$. 
 In other words, multiple SLE processes can only meet at infinity or at their endpoints.

We now consider the events $\{ T<\infty \, , \; I_T = 0 \}$.  The behaviour of the RN-derivative will depend on the singular behaviour of the function $Z[x]$ as two (or more) points come together.  Straightaway we see that if $Z[x]$ has the first type of singular behaviour \eqref{eq:Zsing3}, the limit $x_i \to x_{i{+}1}$ is well defined and the probability that the curve $i$ and $i{+}1$ meet is zero.  In the case where $Z[x]$ is of the second type we have to deal with the singularity.

This singularity represents the fact that while two independent SLEs will meet for the first time at their tips with probability zero, we expect this event will have a finite probability for our multiple SLEs.  Hence the measure $Q$ will be singular with respect to the independent measure $P$ and the singularity indicates this fact.  All is not lost as we do expect to be able to use our RN-derivate to define the measure $Q$ even in this situation, possibly using an extension theorem (Tulcea's for example).  Sadly we are unable to perform this analysis.

If one can make sense of $\lim_{t \to T} Q_t$, then it is very natural extend $Q_t$ to times $t>T$ as suggested in BBK.  Since $Z[x]$ has the product for \eqref{eq:Zsing1}, we can continue the evolution beyond the collision of curves $i$ and $i+1$ by setting $a_t^i = a_t^{i+1} = 0$ and growing $n-2$ curves using the driving functions derived from $Z_1[x]$.  All that we have said so far applies to this new system which will be stopped at some time $T_2$ when (with probability one) the next pair of curves collide at their tips.  After iterating this process, the remaining curves will go to infinity. \\

In the above, we have concentrated on $\kappa_i \le \tfrac 83$.  We expect much of what we have said to extend in principle to $\kappa_i \le 4$ or with appropriate modifications to $\kappa_i < 8$.  However at present we do not have enough control over the RN-derivative to say anything concrete in these cases.

\section{Conclusions}
\label{sec:conclusion}

We have studied a number facets of multiple SLE processes.  We have shown that using  assumptions of absolute continuity, conformal invariance and reparameterisation invariance, one may rederive the result of BBK \cite{BauBerKyt0503} obtained previously by assuming a connection to conformal field theory.

We have shown that absolute continuity implies that the quadratic variation of the driving martingale for multiple SLEs should be related to that of a single SLE.
We also showed how conformal invariance and reparameterisation invariance imply the form of the multiple SLE driving functions proposed by BBK.   As a corollary of this analysis we saw that the $\kappa$ parameters of each component multiple SLE should be related $\kappa_i = \kappa_j$ or $\kappa_i = \tfrac{16}{\kappa_j}$, or in other words, the component SLEs should have the same central charge.  This provides a probability theory realisation of both $\phi_{1,2}$ and $\phi_{2,1}$ fields found in conformal field theory.
In the final section, we studied reparameterisation invariance by conditioning $n$-independent SLEs.  Moreover, we used the associated Radon-Nykodym derivative to study properties of multiple SLEs in the regime $\kappa_i \le \tfrac 83$.

Although we believe the general picture is clear, as always with probability there is a great deal of devil in the detail, an in particular, our arguments involving the RN-derivative are far from complete.  As a final point, we have discussed many sufficient conditions.  It would be interesting to see how much of this structure is actually necessary.

\section{Acknowledgments}

The Author is grateful for conversations with Roland Friedrich, Denis Bernard, John Cardy and Gregory Lawler.  This work was supported by the EUCLID Network, contract number HPRN-CT-2002-00325.

\appsection

In this appendix we discuss the consequences of choosing a curve parameterisation such that $a_t^i = 1$ for all $i$.  From equation \eqref{eq:afromc} we note,
\begin{gather}
  \hcap[ K_t ] = \int_0^t \sum_{i=1}^n a_s^i ds = nt \; , \\
  a_t^i = {H_t^i}'(w_t^i)^2 c_t^i = 1 \; , 
  \qquad
  \hcap[ K_t^i ]= \int_0^t c_s^i ds  = \int_0^t \frac{1}{{H_s^i}'(w_s^i)^2} ds \; . \label{eq:hcapusingH}
\end{gather}
It also follows from \eqref{eq:afromc} that,
\begin{align}
  \hcap[ K_t^i \cup K_t^j ] &=  \int_0^t\!\! \left[ {H_{j,s}^i}'(w_s^i)^2 c_s^i + {H_{i,s}^j}'(w_s^j)^2 c_s^j \right] ds \notag \\ 
  &= \! \int_0^t \!\! \left[ \frac{1}{{H_{s}^{ij}}'(H_{j,s}^i(w_s^i))^2} + \frac{1}{{H_{s}^{ij}}'(H_{i,s}^j(w_s^j))^2} \right] ds \; , \label{eq:hcapfor2} 
\end{align}
where we have introduced new maps, $H_{j,t}^i : \mH / g_t^i(K_t^j) \to \mH$ and $H^{ij}_t = H_t^i \circ {H_{j,t}^i}^{-1}$.  The idea behind the notation is as follows:  let $H_{i_1, \ldots i_m,t} : \mH / \cup_{r=1}^m K_t^{i_r} \to \mH$ then we define $H_{j_1 \ldots j_p,t}^{i_1 \ldots i_m} : \mH / H_{i_1, \ldots i_m,t}( \cup_{r=1}^p K_t^{j_r} ) \to \mH$.  To reduce indices we also define $H^{i_1,\ldots,i_m}_t = H_{j_1 \ldots j_p,t}^{i_1 \ldots i_m}$ when the set $\{ j_r \}_{r=1}^p$ contains all indices not included in the set $\{ i_r \}_{r=1}^m$, in other words $G_t = H^{i_1,\ldots,i_m}_t \circ H_{i_1, \ldots i_m,t}$.  In particular $H_{i,t} = g_t^i$ and, although we will never use it, $H_t = H_{1 \ldots n,t} = G_t$. \\

\newcommand{\mb}[2]{\makebox(0,0)[#1]{$\ss{#2}$}}

\begin{figure}
\refstepcounter{figure}
\label{fig:Hhat}
\begin{center}
\begin{tabular}{c}
\setlength{\unitlength}{1cm}
\begin{picture}(14.5,9)
  \put(0,6){
     \epsfxsize 6.5\unitlength
     \epsfbox{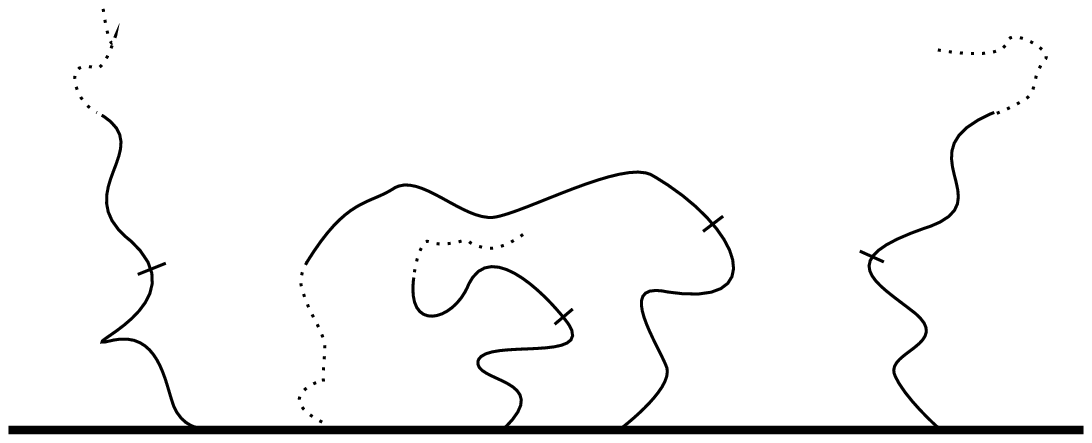}}
  \put(8,6){
     \epsfxsize 6.5\unitlength
     \epsfbox{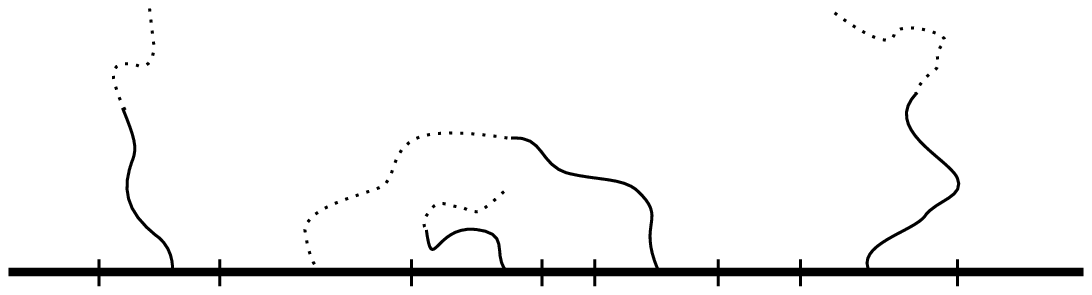}}
  \put(0,3){
     \epsfxsize 6.5\unitlength
     \epsfbox{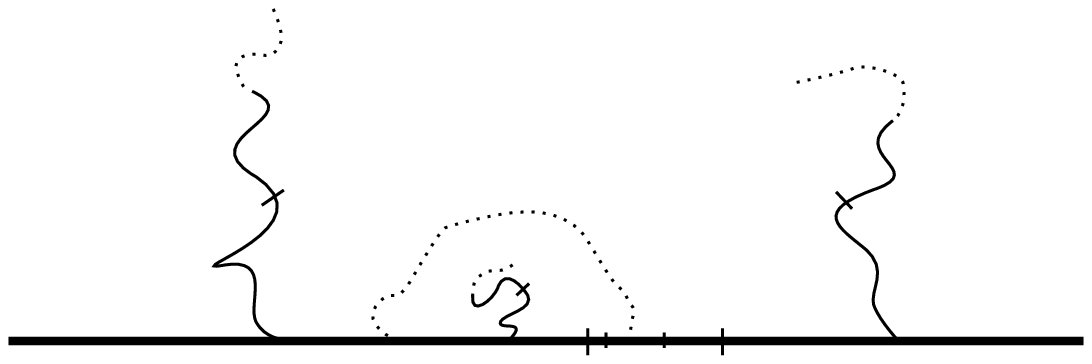}}
  \put(8,3){
     \epsfxsize 6.5\unitlength
     \epsfbox{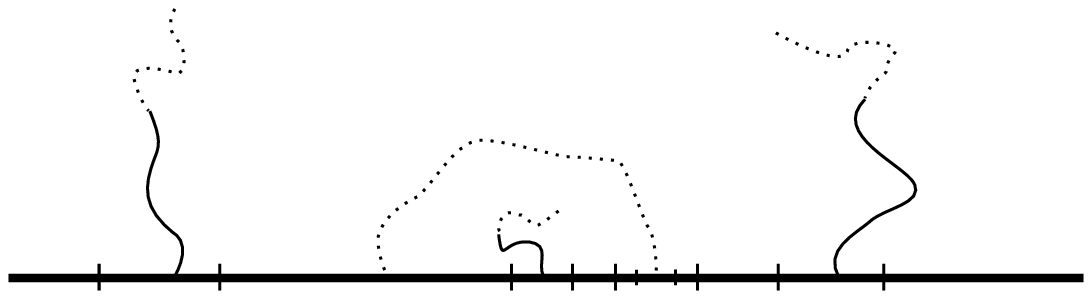}}
  \put(4.5,0){
     \epsfxsize 6.5\unitlength
     \epsfbox{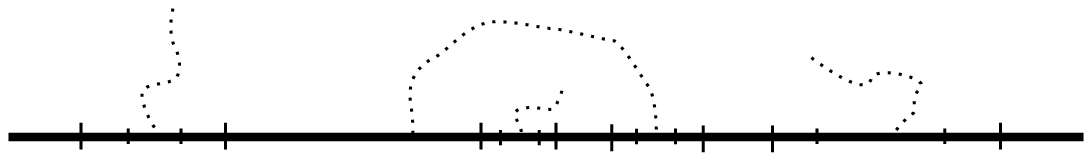}}
  \put(2.55,7.4){
     \epsfysize 0.3\unitlength
     \epsfbox{figures/arrow_down.eps}}
  \put(2.4,6.3){
     \epsfysize 0.3\unitlength
     \epsfbox{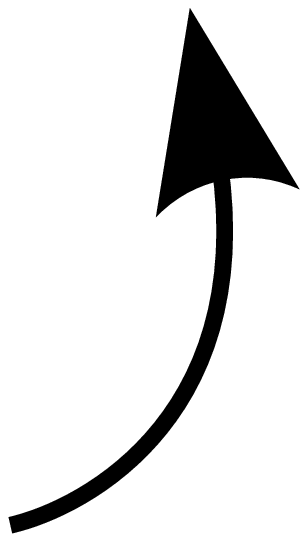}}
  \put(6.5,6.6){
     \epsfxsize 1\unitlength
     \epsfbox{figures/arrow_s_left.eps}}
  \put(6.5,3.6){
     \epsfxsize 1\unitlength
     \epsfbox{figures/arrow_s_left.eps}}
  \put(2.75,4.75){
     \epsfysize 1\unitlength
     \epsfbox{figures/arrow_s_down.eps}}
  \put(10.75,4.75){
     \epsfysize 1\unitlength
     \epsfbox{figures/arrow_s_down.eps}}
  \put(9.85,1.5){
     \epsfysize 1\unitlength
     \epsfbox{figures/arrow_s_downright.eps}}
  \put(3.85,1.5){
     \epsfysize 1\unitlength
     \epsfbox{figures/arrow_s_downleft.eps}}
  \put(2.25,7.8){ \makebox(0,0)[l]{$\gamma^i$}}
  \put(2.05,6.4){ \makebox(0,0)[l]{$\gamma^j$}}
  \put(6.75,7){ \makebox(0,0)[l]{$G_s$}}
  \put(3,5.25){ \makebox(0,0)[l]{$g_t^i$}}
  \put(11,5.25){ \makebox(0,0)[l]{$\hat{g}_{s;t}^i$}}
  \put(4.5,2.25){ \makebox(0,0)[l]{$H_t^i$}}
  \put(10.5,1.75){ \makebox(0,0)[l]{$\hat{H}_{s;t}^i$}}
  \put(6.7,4){ \makebox(0,0)[l]{$\Psi_{s;t}$}}
\end{picture}
\vspace{0.5cm} \\
\parbox{12cm}{
\small\raggedright
{\em \begin{center} Figure \ref{fig:Hhat} : A diagram to represent the maps $\hat{H}_{s;t}$ and $\Psi_{s;t}$.\\ 
In the top left figure, the solid lines represent the curves up to time t, the dashes on the curves represent the location of the tips at time s, and the dotted lines represent the curves from time t until time T. \end{center}}}
\end{tabular}
\end{center}
\end{figure}

We will now show that if $a_t^i = 1$, then one curve cannot enclose
another (ie. cannot disconnect the tip of a curve from infinity).  
Let us assume this is false and that a curve $i$ encloses a
curve $j$ for the first time at time $T$.  Furthermore, let us assume
the curve $i$ does 
not meet the tip of another curve at time $T$ (we will deal with this
case in a moment) and that the curve $i$ itself is not enclosed by
another curve at time $T$.  Since $i$ is enclosing $j$, at $T$ it must hit
either the real axis or another curve $k$ at a
point $\gamma_r^k=\gamma_T^i$ for some time $r < T$ (possibly $k=j$, but not $k=i$).  In the second
case we may choose a time $s$, $r<s<T$ and consider the curves
$\hat{\gamma}_t^\ell = G_s(\gamma_t^\ell)$ instead.  Hence without loss of
generality, we may assume the curve $\gamma_t^i$ collides with the
real axis at $T$.  Furthermore, (with a suitable choice of $s$) we
assume we can arrange that there exists a small neighbourhood around the point
$\gamma_T^i$ which contains only the curve $\gamma_T^i$\footnote{This assumption
 excludes some curves.}.  Under these assumptions, the object
${H_t^i}'(w_t^i)$ has a well defined limit as $t \to T$ and
${H_T^i}'(w_T^i)>0$.

Since $i$ collides with the real axis,
\begin{align}
  \hcap[ K_T^i ] = \hcap[ K_T^i \cup K_T^j ] \; .
\end{align}
Moreover because $0 \le {H^{ij}_t}'(z) \le 1$ we see from \eqref{eq:hcapfor2} that $\hcap[ K_T^i \cup K_T^j ] \ge 2T$ and hence from \eqref{eq:hcapusingH}, there exists a time $t_0$,  $0 \le t_0 \le T$ such that,
\begin{align}
  \frac{1}{{H_{t_0}^i}'(w_{t_0}^i)^2} \ge \frac{\hcap[K_T^i]}{T} \ge 2 \; .
\end{align}
Let us introduce a time $s$, $0 < s < T$, and maps $\hat{H}^{i_1 \ldots i_m}_{s;j_1,\ldots,j_p,t}$ for $t$, $s \le t \le T$ defined as $H$ above but for the sets $\hat{K}_t^i = G_s(K_t^i)$ (an example of such a map is represented in figure \ref{fig:Hhat}).
It is not difficult to check that we can also apply our argument to $\hat{H}_{s;t}^i$ and so obtain that there exists a $t_1$,  $s \le t_1 \le T$ such that,
\begin{align}
  \frac{1}{{ {\hat{H}}_{s;t_1}^i }{}'(w_{t_1}^i)^2 } \ge 2 \; .
\end{align} 
To connect $\hat{H}_{s,t}^i$ to $H_t^i$, we define $\Psi_{s;t}= \hat{g}_{s;t}^i \circ G_s \circ {g_t^i}^{-1}$ (see figure \ref{fig:Hhat}) so $H_t^i = \hat{H}_{s;t}^i \circ \Psi_{s;t}$.  
By setting $s_n = \tfrac 12 (T+t_{n-1})$ we construct a sequence $t_n \to T$,
\begin{align}
  \frac{\Psi_{s_n;t_n}'(w_{t_n}^i)^2}{{H_{t_n}^i}'(w_{t_n}^i)^2} \ge  2  \; .
  \label{eq:inequality}
\end{align}
Now  $\lim_{n \to \infty} {H_{t_n}^i}'(w_{t_n}^i)^2 = \lim_{n \to
  \infty} {H_{t_n}^{ij}}'( H_{j,t_n}^i(w_{t_n}^i) )
{H_{j,{t_n}}^i}'(w_{t_n}^i) =  {H_T^{ij}}'( w_T^i )>0$ using our
initial assumptions.  
On the other hand $\Psi_{s;t}(z)$ as a function of $s$ is the multiple Loewner evolution for $g_t^i(K_s)$ and as such, the Loewner equation shows $\Psi'_{s;t}(w_t^i)$ is a continuous decreasing function of $s$, $\Psi'_{s_n;t_n}(w_{t_n}^i) \le  \Psi_{s_m;t_n}'(w_{t_n}^i)$ for $m < n$ and so,
\begin{align}
  \lim_{n \to \infty} \Psi_{s_n;t_n}'(w_{t_n}^i) \le  \lim_{n \to \infty} \Psi_{s_m;t_n}'(w_{t_n}^i) = \Psi_{s_m;T}'(w_T^i) \; \qquad \text{for all}\;\; m \; .
\end{align}
Since $\lim_{m \to \infty} \Psi_{s_m;T}'(w_T^i) = {H_T^{ij}}'( w_T^i )$ we see the limit of the LHS of \eqref{eq:inequality} is bounded above by $1$ and we obtain our contradiction. \\

We can extend these arguments to show that if $a_t^i=1$, two curves
joining at the tip cannot enclose one or more (growing) curves.  Let
us assume that curves $i$ and $j$ meet at time $T$ enclosing a curve
$k$.  Proceeding as before, a formula similar to \eqref{eq:hcapfor2}
shows that $\hcap[ K_T^i \cup K_T^j \cup K_T^k ] \ge 3T$.  However,
since $\lim_{t \to T} \hcap[ K_t^i \cup K_t^j ] = \hcap[ K_T^i \cup
K_T^j \cup K_T^k ]$ we know there exists a time $0 \le t_0 \le T$ such
that, 
\begin{align}
  \frac{1}{{H_{t_0}^{ij}}'( H_{j,t_0}^i(w_{t_0}^i) )^2 } +   \frac{1}{{H_{t_0}^{ij}}'( H_{i,t_0}^j(w_{t_0}^j) )^2 } \ge 3 \; .
\end{align}
This relation is also true for $\hat{H}_{s;t_1}^{ij}$, $s \le t_1 \le T$ and hence by setting $s_n = \tfrac 12 (T+t_{n-1})$ we construct a sequence $t_n \to T$,
\begin{align}
   &\frac{1}{{\hat{H}_{s_n;t_n}^{ij}}{}'( \hat{H}_{s_n;j,t_n}^i(w_{t_n}^i) )^2 } +   \frac{1}{{\hat{H}_{s_n;t_n}^{ij}}{}'( \hat{H}_{s_n;i,t_n}^j(w_{t_n}^j) )^2 } \notag \\
&=  \frac{\Phi_{s_n;t_n}'(H_{j,t_n}^i(w_{t_n}^i))^2 }{{H_{t_n}^{ij}}'( H_{j,t_n}^i(w_{t_n}^i) )^2 } +   \frac{\Phi_{s_n;t_n}'(H_{i,t_n}^j(w_{t_n}^j))^2 }{{H_{t_n}^{ij}}'( H_{i,t_n}^j(w_{t_n}^j) )^2 }
\ge 3 \; ,
\end{align}
where $H_t^{ij} = \hat{H}_{s;t}^{ij} \circ \Phi_{s;t}$.  
One then proceeds as before to check that $\lim_{n \to \infty} \Phi_{s_n;t_n}'(H_{i,t_n}^j(w_{t_n}^j)) \le \lim_{n \to \infty} {H_{t_n}^{ij}}'( H_{i,t_n}^j(w_{t_n}^j)$ and hence the LHS is bounded above by $2$ giving the contradiction.

\bibliographystyle{utcaps}
\bibliography{references}

\providecommand{\href}[2]{#2}\begingroup\raggedright\begin{thebibliography}{10}

\bibitem{Sch9904}
O.~Schramm, ``Scaling limit of loop-erased random walks and uniform spanning
  trees,'' {\em Israel J. Math.} {\bf 118} (2000) 221--288,
\href{http://www.arXiv.org/abs/math/9904022}{{\tt math/9904022}}.

\bibitem{RohSch0106}
S.~Rohde and O.~Schramm, ``Basic properties of SLE,''
\href{http://www.arXiv.org/abs/math/0106036}{{\tt math/0106036}}.

\bibitem{Wer0303}
W.~Werner, ``Random planar curves and Schramm-Loewner evolutions,''
\href{http://www.arXiv.org/abs/math/0303354}{{\tt math/0303354}}.

\bibitem{KagNie0312}
W.~Kager and B.~Nienhuis, ``A Guide to Stochastic Loewner Evolution and its
  Applications,''
\href{http://www.arXiv.org/abs/math-ph/0312056}{{\tt math-ph/0312056}}.

\bibitem{book:Law}
G.~Lawler, {\em Conformally Invariant Processes in the Plane}.
\newblock AMS, 1st~ed., 2005.

\bibitem{BauBer0210}
M.~Bauer and D.~Bernard, ``Conformal field theories of stochastic Loewner
  evolutions,'' {\em Commun. Math. Phys.} {\bf 239} (2003) 493--521,
\href{http://www.arXiv.org/abs/hep-th/0210015}{{\tt hep-th/0210015}}.

\bibitem{FriWer0301}
R.~Friedrich and W.~Werner, ``Conformal restriction, highest-weight
  representations and SLE,''
\href{http://www.arXiv.org/abs/math/0301018}{{\tt math/0301018}}.

\bibitem{BauBer0305}
M.~Bauer and D.~Bernard, ``Conformal transformations and the SLE partition
  function martingale,'' {\em Annales Henri Poincare} {\bf 5} (2004) 289--326,
\href{http://www.arXiv.org/abs/math-ph/0305061}{{\tt math-ph/0305061}}.

\bibitem{BauBer0401}
M.~Bauer and D.~Bernard, ``SLE, CFT and zig-zag probabilities,''
\href{http://www.arXiv.org/abs/math-ph/0401019}{{\tt math-ph/0401019}}.

\bibitem{Car0310}
J.~Cardy, ``Calogero-Sutherland model and bulk-boundary correlations in
  conformal field theory,'' {\em Phys. Lett.} {\bf B582} (2004) 121--126,
\href{http://www.arXiv.org/abs/hep-th/0310291}{{\tt hep-th/0310291}}.

\bibitem{Dub0411}
J.~Dub\'edat, ``Some remarks on commutation relations for SLE,'' (2004)
\href{http://www.arXiv.org/abs/math/0411299}{{\tt math/0411299}}.

\bibitem{BauBerKyt0503}
M.~Bauer, D.~Bernard, and K.~Kytola, ``Multiple Schra-Loewner Evolutions and
  Statistical Mechanics Martingales,''
\href{http://www.arXiv.org/abs/math-ph/0503024}{{\tt math-ph/0503024}}.

\bibitem{book:Yor}
D.~Revuz and M.~Yor, {\em Continuous Martingales and Brownian Motion}.
\newblock Springer, 3rd~ed., 2001.

\bibitem{LawSchWer0209}
G.~Lawler, O.~Schramm, and W.~Werner, ``Conformal restriction: the chordal
  case,'' {\bf 1}
\href{http://www.arXiv.org/abs/math/0209343}{{\tt math/0209343}}.

\bibitem{Dub0303}
J.~Dub\'edat, ``SLE(kappa,rho) martingales and duality,'' {\em Ann. Prob.} {\bf
  33, no.1} (2005) 233--243,
\href{http://www.arXiv.org/abs/math/0303128}{{\tt math/0303128}}.

\bibitem{Kyt0504}
K.~Kytola, ``On conformal field theory of SLE(kappa,rho),''
\href{http://www.arXiv.org/abs/math-ph/0504057}{{\tt math-ph/0504057}}.

\bibitem{BetGruLudWie0503}
E.~Bettelheim, I.~Gruzberg, A.~W.~W. Ludwig, and P.~Wiegmann, ``Stochastic
  Loewner evolution for conformal field theories with Lie-group symmetries,''
\href{http://www.arXiv.org/abs/hep-th/0503013}{{\tt hep-th/0503013}}.

\bibitem{Dub0507}
J.~Dub\'edat, ``Euler integrals for commuting SLEs,'' (2005)
\href{http://www.arXiv.org/abs/math/0507276}{{\tt math/0507276}}.

\bibitem{BelPolZam84}
A.~A. Belavin, A.~M. Polyakov, and A.~B. Zamolodchikov, ``Infinite conformal
  symmetry in two-dimensional quantum field theory,'' {\em Nucl. Phys.} {\bf
  B241} (1984)
333--380.

\bibitem{DotFat84}
V.~S. Dotsenko and V.~A. Fateev, ``Conformal algebra and multipoint correlation
  functions in 2D statistical models,'' {\em Nucl. Phys.} {\bf B240} (1984)
312.

\end{thebibliography}\endgroup

\end{document}